\documentclass{article}

\usepackage{amssymb,graphicx}

\begin{document}


\title{\vspace{-2cm}Urban sprawl and evolution of accessibility profiles in Chinese cities}


\author{J. Raimbault$^{1,2,3,\ast}$\\
$^{1}$ Center for Advanced Spatial Analysis, University College London\\
$^{2}$ UPS CNRS 3611 ISC-PIF\\
$^{3}$ UMR CNRS 8504 G{\'e}ographie-cit{\'e}s\medskip\\
$\ast$ \texttt{juste.raimbault@polytechnique.edu}
}

\date{}

\maketitle

\begin{abstract}
The development of public transportation networks and associated transit oriented development policies are efficient tools to mitigate urban sprawl and its negative environmental impacts, especially in terms of commuting emissions. We study in this paper the trajectories in terms of sprawl and public transport accessibility of the nine largest Chinese megacities, from 2000 to 2030 with projected transportation networks and populations. Using the radial profile methodology, we compare the evolution of population distribution with the evolution of accessibility profiles. We show a rebalancing of access in time, suggesting the efficiency of transport network growth regarding transit-oriented development objectives.
\end{abstract}

\section{Introduction}

\subsection{Urban sustainability and public transport accessibility}

Sustainable Development Goals include both to reach very low greenhouse gases emissions (goal 13), build sustainable cities (goal 11), while maintaining economic activity and fostering innovation (goals 8 and 9) \cite{griggs2014integrated}. While key issues of urban sustainability are linked to energy production and a circular economy of resources \cite{kennedy2012sustainable}, low carbon mobility is essential to ensure an equity of access to economic opportunities with limited emissions. Planning and management of public transport, in particular offering a high level of public transport accessibility within densely populated zones of urban areas, is in that context an important aspect of urban sustainability \cite{hayashi2004urban}.

\subsection{Urban sprawl and transit-oriented development}

Urban form, in the sense of the spatial distribution of population, can have a significant impact on accessibility. Sprawled cities can be opposed to more compact ones \cite{tsai2005quantifying} in which transit lines reach more users. \cite{le2012urban} showed for European cities that sprawled or polycentric cities were effectively less sustainable in terms of energy-efficient mobility. Countering urban sprawl through planning implies thus some careful planning of transit lines. Transit-oriented development proposes to couple public transport with dense areas around stations \cite{dittmar2012new}. This corresponds in some sense to fostering the co-evolution between land-use and the transport network \cite{raimbault2021characterising}. This approach has been used intensively in recent years in Asian cities, in particular in China with a quick development of transit networks \cite{raimbault2018evolving}. We will focus on this case study in the rest of this paper.

\subsection{Homothetic scaling and accessibility profiles}

The size and spatial structure of cities are tightly related, as it was shown by \cite{lemoy2020evidence} that full population density radial profiles follow a scaling relationship with population. This was coined as ``homethetic scaling'', as all cities appear then to be scaled versions of another. This property can be derived from the Alonso urban monocentric model \cite{delloye2020alonso}. It furthermore holds for a variety of land-uses \cite{lemoy2021radial}.

Travel times within large functional urban areas similarly follow some radial scaling patterns: \cite{mennicken2019internal} show that congestions leads to higher transport times the closer to the center, and provide empirical evidence for homothetic scaling of transport access. As the radial profile methodology is tightly linked to the monocentric model and thus to sprawl, we will use it to study the link between urban sprawl and the distribution of public transport accessibility.

\subsection{Proposed approach}

We propose in this paper to study the interaction between transport and land-use, using the homothetic radial scaling methodology. Comparing population profiles with accessibility profiles informs on which proportion of population has a reasonable level of accessibility, and in particular if peripheral areas are well served by the network.

We study nine Chinese megacities which have since around 2000 witnessed a very quick development of high coverage subway networks. The cities are Beijing, Chengdu, Chongqing, Guangzhou, Nanjing, Shanghai, Shenzhen, Wuhan, Xian. For example, Shenzhen planned for each district to be within 500m reach of a station \cite{ng2011strategic}. Most of the networks have in 2021 already reached a mature stage \cite{raimbault2018evolving}, and the remaining developments until 2030 will complete these. 

The rest of this paper is organised as follows: we first detail data and measures used; we then present accessibility maps, and compare radial profiles for population and accessibility; we finally discuss potential applications and future developments.

\section{Methods}

\subsection{Data}

Population data is a 1km raster, provided by the Global Human Settlement Layer database \cite{florczyk2019ghsl}. Populations are available for years 1990, 2000, 2015. For the extent of urban areas, we use functional areas from the same database. Public transport networks come from a vector database, built by \cite{raimbault2018evolving}, and providing subway lines and stations with their opening dates. Such a dynamical database allows studying the evolution of network measures.

\subsection{Accessibility measures}

Numerous measures and interpretations exist to measure accessibility \cite{levinson2020towards}. As we do not use projected population, only the 2015 raster is available for the 2030 network. Therefore, using population weighted measures does not make sense to study accessibility dynamics. We principally use a gravity-like measure of access, similar to some closeness centrality, given for an area $i$ by

\begin{equation}
\label{eq:access}
Z_i = \frac{1}{N} \sum_j \exp\left(-t_{ij} / t_0\right)
\end{equation}

with $N$ areas and $t_{ij}$ travel time through the subway network, and $t_0$ a characteristic time corresponding to the typical commute time (which we take in practice as $t_0 = 1h$).

A destination population weighted measure would include in the summation the share of population for each area. This measure can also be weighted at the origin, or by some amenity at destination.

To compute travel time, we assume closest station is reached with a speed of 25km/h (car in congested cities), and that subway lines have an average speed of 50km/h. We do not consider a multi-modal network for the sake of simplicity, as we focus on public transport times only. We do not consider exchanges times also to simplify the computations, and use a simple shortest path routing. Finally, we assume a high level of service and that accessibility is computed at peak time, and therefore neglect waiting times.

\section{Results}

\subsection{Accessibility maps}

\begin{figure}
	\includegraphics[width=\linewidth]{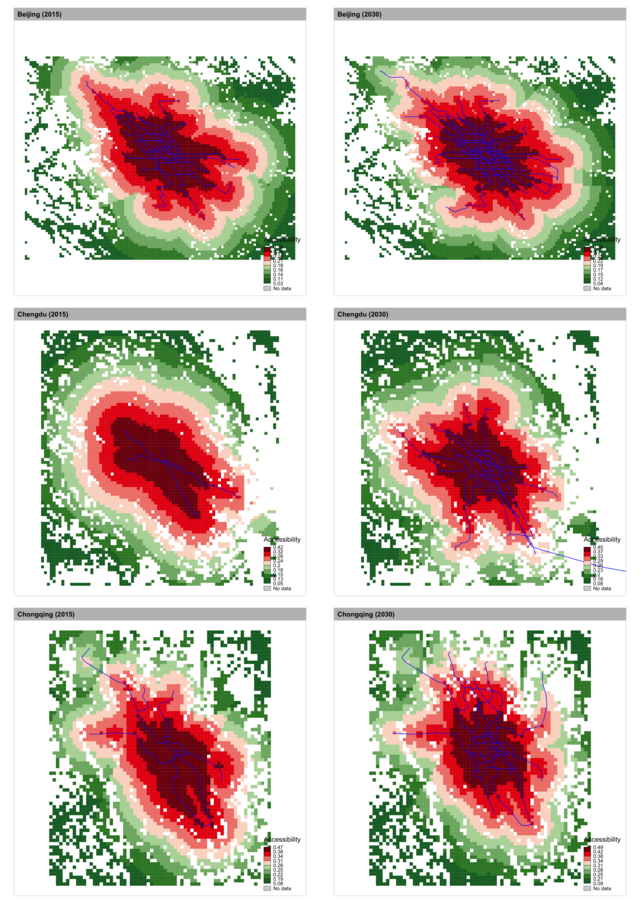}
	\caption{Temporal accessibility maps computed with equation~\ref{eq:access}, in 2015 and 2030, for Beijing, Chengdu and Chongqing. Colors are obtained with equal size groups. We show the subway network in blue.\label{fig:fig1}}
\end{figure}

\begin{figure}
	\includegraphics[width=\linewidth]{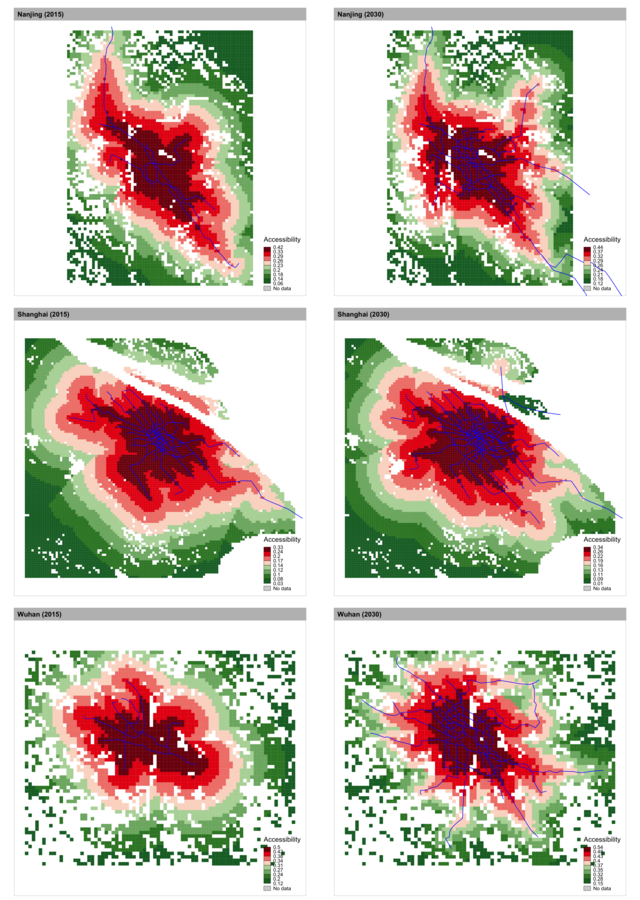}
	\caption{Temporal accessibility maps for Nanjing, Shanghai and Wuhan, in 2015 and 2030.\label{fig:fig2}}
\end{figure}

We show accessibility maps for all 1km raster cells within the functional urban areas, for Beijing, Chengdu and Chongqing in Fig.~\ref{fig:fig1}, and for Nanjing, Shanghai and Wuhan in Fig.~\ref{fig:fig2}. The three remaining cities are not shown here but present very similar patterns.

The most striking feature is the reinforcement of access as a monocentric feature, in the sense that most of the center of the area becomes highly and mostly equally accessible. Except for Beijing and Shanghai which already had in 2015 a well developed network, all other cities strongly densify transit lines in the core area, ensuring a balanced accessibility within it. Beyond the center, radial transit lines distribute access in a transit-oriented development manner.

\subsection{Radial profiles}

\begin{figure}
	\includegraphics[width=\linewidth]{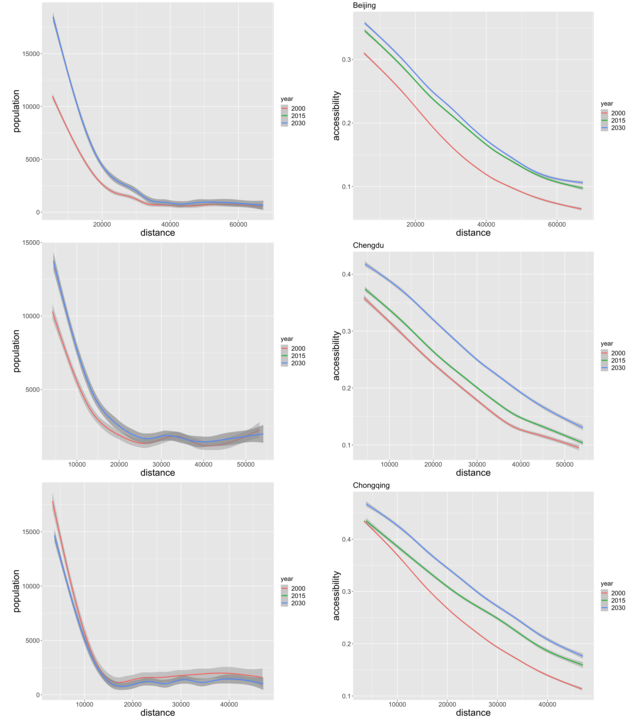}
	\caption{Population (left) and temporal accessibility (right) radial profiles, in 2000, 2015 and 2030 (curve color), for Beijing, Chengdu and Chongqing. Curves show the average of cells within the corresponding distance bin, with confidence interval in grey.\label{fig:fig3}}
\end{figure}

\begin{figure}
	\includegraphics[width=\linewidth]{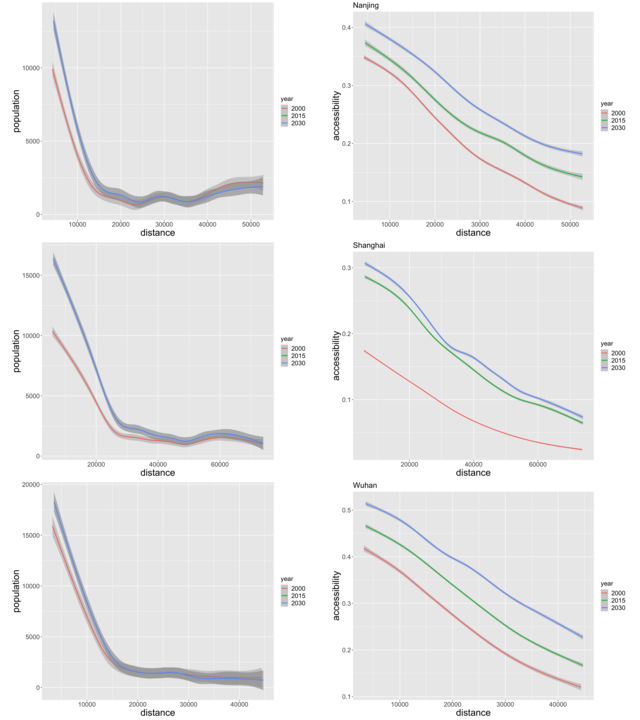}
	\caption{Population and temporal accessibility radial profiles, in 2000, 2015 and 2030 (curve color), for Nanjing, Shanghai and Wuhan.\label{fig:fig4}}
\end{figure}

We compute radial profiles, from the topological center of the functional urban area (centroid), by binning raster cells into 5\% quantiles. We consider the average population and temporal accessibility for all cells within a given bin. We compute separately profiles for population and for accessibility, as we do not consider population-weighted accessibility measures. With this small number of cities, we do not consider potential scaling properties and homothetic reduction of these profiles as done in the original methodology by \cite{lemoy2020evidence}.

The profiles are shown in Fig.~\ref{fig:fig3} for Beijing, Chengdu and Chongqing, and in Fig.~\ref{fig:fig4} for Nanjing, Shanghai and Wuhan. All population profiles reach a minimum within a distance of around 20km from the center, which corresponds to the approximate size of the densely populated urban core. For some cities like Chengdu and Nanjing, a slight increase at large distances may be a signature of distant secondary centers. The accessibility profiles show an evolution with translated profiles across the years. In 2030, at the center, most cities have their accessibility increased by 0.1, corresponding to around 30\% of relative increase, which is significant. The translation of the curve corresponds to a strong increase in access equity: for example, considering Beijing, the level of the center in 2000 is reached by cells at 18km from the center in 2030. Looking at the corresponding population, a large majority of it is thus covered, as population profiles become low at around 20km. The highest increase is for Shanghai, for which the center value in 2000 is reached at 40km in 2030. The center value is close to be doubled.

\section{Discussion}

The main feature of transit-oriented development is to ensure that public transport, and most preferably heavy rail transit, become the privileged transport mode, at least in newly developed areas. This can be done using density around stations, but also efficient local transportation to reach the station. A more systematic and efficient view is to develop a very dense network within reach of most populated areas within the urban core. It seems from our results that this strategy was considered by Chinese authorities, as most of population within the studied areas will be within close reach of a subway station in 2030. This is confirmed by the profiles, and more particularly by comparing population and accessibility radial profiles. At this scale of a single urban center, the monocentric model is in some sense still in place but the very high network density ensures some equity in accessibility, as all areas are above a high value.

At other scales, such as polycentric mega-city regions \cite{raimbault2021introducing}, different questions and planning issues can be raised, as low carbon mode of transport will be high speed rail for example \cite{wang2013spatial}. An efficient articulation between the local, intermediate and national scale, is necessary to foster urban sustainability at these larger scales, and to ensure that no detrimental effects emerge for some urban centers \cite{wang2018high}.

Our study remained simple and could be extended in several directions: using multi-modal accessibility and a more detailed description of the public transport service, including off-peak hours and exchange times. Then, looking at more general measures of accessibility would also be useful, also detailed amenity and land-use data is more difficult to obtain. Finally, generalising beyond subway networks would allow having a larger sample of cities, and thus to study the homothetic scaling properties of accessibility profiles.

To conclude, we have studied population and temporal accessibility radial profiles, for nine Chinese megacities, between 2000 and 2030. We strengthen the conclusions of \cite{raimbault2018evolving}, which had found a high increase in transport equity with the developing transit networks, as we establish the correspondance between population and accessibility profiles, which shows how most of the population benefits with the final network from a very high accessibility level. This suggests that Chinese cities are on a good path to reach urban sustainability, at least for the low-carbon mobility dimension. Worldwide comparison such as \cite{wu2021urban} would be useful to understand how to transfer these urban successes to other regions of the world.


\begin{thebibliography}{}

\bibitem[Delloye et~al., 2020]{delloye2020alonso}
Delloye, J., Lemoy, R., and Caruso, G. (2020).
\newblock Alonso and the scaling of urban profiles.
\newblock {\em Geographical Analysis}, 52(2):127--154.

\bibitem[Dittmar and Ohland, 2012]{dittmar2012new}
Dittmar, H. and Ohland, G. (2012).
\newblock {\em The new transit town: Best practices in transit-oriented
  development}.
\newblock Island Press.

\bibitem[Florczyk et~al., 2019]{florczyk2019ghsl}
Florczyk, A.~J., Corbane, C., Ehrlich, D., Freire, S., Kemper, T., Maffenini,
  L., Melchiorri, M., Pesaresi, M., Politis, P., Schiavina, M., et~al. (2019).
\newblock Ghsl data package 2019.
\newblock {\em Luxembourg, EUR}, 29788(10.2760):290498.

\bibitem[Griggs et~al., 2014]{griggs2014integrated}
Griggs, D., Smith, M.~S., Rockstr{\"o}m, J., {\"O}hman, M.~C., Gaffney, O.,
  Glaser, G., Kanie, N., Noble, I., Steffen, W., and Shyamsundar, P. (2014).
\newblock An integrated framework for sustainable development goals.
\newblock {\em Ecology and society}, 19(4).

\bibitem[Hayashi et~al., 2004]{hayashi2004urban}
Hayashi, Y., Doi, K., Yagishita, M., and Kuwata, M. (2004).
\newblock Urban transport sustainability.
\newblock {\em European Journal of Transport and Infrastructure Research},
  4(1).

\bibitem[Kennedy et~al., 2012]{kennedy2012sustainable}
Kennedy, C., Baker, L., Dhakal, S., Ramaswami, A., et~al. (2012).
\newblock Sustainable urban systems.
\newblock {\em Journal of Industrial Ecology}, 16(6):775--974.

\bibitem[Le~N{\'e}chet, 2012]{le2012urban}
Le~N{\'e}chet, F. (2012).
\newblock Urban spatial structure, daily mobility and energy consumption: a
  study of 34 european cities.
\newblock {\em Cybergeo: European Journal of Geography}.

\bibitem[Lemoy and Caruso, 2020]{lemoy2020evidence}
Lemoy, R. and Caruso, G. (2020).
\newblock Evidence for the homothetic scaling of urban forms.
\newblock {\em Environment and Planning B: Urban Analytics and City Science},
  47(5):870--888.

\bibitem[Lemoy and Caruso, 2021]{lemoy2021radial}
Lemoy, R. and Caruso, G. (2021).
\newblock Radial analysis and scaling of urban land use.
\newblock {\em Scientific reports}, 11(1):1--8.

\bibitem[Levinson and Wu, 2020]{levinson2020towards}
Levinson, D. and Wu, H. (2020).
\newblock Towards a general theory of access.
\newblock {\em Journal of Transport and Land Use}, 13(1):129--158.

\bibitem[Mennicken et~al., 2019]{mennicken2019internal}
Mennicken, E., Caruso, G., and Lemoy, R. (2019).
\newblock Internal radial profiles of road transport times in european cities.
\newblock {\em ECTQG 2019}.

\bibitem[Ng, 2011]{ng2011strategic}
Ng, M.~K. (2011).
\newblock Strategic planning of china's first special economic zone: Shenzhen
  city master plan (2010--2020).

\bibitem[Raimbault, 2018]{raimbault2018evolving}
Raimbault, J. (2018).
\newblock Evolving accessibility landscapes: mutations of transportation
  networks in china.
\newblock {\em arXiv preprint arXiv:1812.01473}.

\bibitem[Raimbault, 2021]{raimbault2021characterising}
Raimbault, J. (2021).
\newblock Characterising and modeling the co-evolution of transportation
  networks and territories.
\newblock {\em arXiv preprint arXiv:2110.15950}.

\bibitem[Raimbault and Le~N{\'e}chet, 2021]{raimbault2021introducing}
Raimbault, J. and Le~N{\'e}chet, F. (2021).
\newblock Introducing endogenous transport provision in a luti model to explore
  polycentric governance systems.
\newblock {\em Journal of Transport Geography}, 94:103115.

\bibitem[Tsai, 2005]{tsai2005quantifying}
Tsai, Y.-H. (2005).
\newblock Quantifying urban form: compactness versus' sprawl'.
\newblock {\em Urban studies}, 42(1):141--161.

\bibitem[Wang et~al., 2013]{wang2013spatial}
Wang, J.~J., Xu, J., and He, J. (2013).
\newblock Spatial impacts of high-speed railways in china: a total-travel-time
  approach.
\newblock {\em Environment and planning A}, 45(9):2261--2280.

\bibitem[Wang and Duan, 2018]{wang2018high}
Wang, L. and Duan, X. (2018).
\newblock High-speed rail network development and winner and loser cities in
  megaregions: The case study of yangtze river delta, china.
\newblock {\em Cities}, 83:71--82.

\bibitem[Wu et~al., 2021]{wu2021urban}
Wu, H., Avner, P., Boisjoly, G., Braga, C.~K., El-Geneidy, A., Huang, J.,
  Kerzhner, T., Murphy, B., Niedzielski, M.~A., Pereira, R.~H., et~al. (2021).
\newblock Urban access across the globe: an international comparison of
  different transport modes.
\newblock {\em npj Urban Sustainability}, 1(1):1--9.

\end{thebibliography}



\end{document}